\documentclass[prl,twocolumn,superscriptaddress,showpacs]{revtex4}
\usepackage{graphicx}
\usepackage{epsfig}
\usepackage{amssymb,amsmath,amsfonts,hyperref}
\usepackage{wasysym}
\usepackage{latexsym}
\usepackage{eucal}
\usepackage{verbatim}
\usepackage[normalem]{ulem}

\usepackage{color}

\newcommand{\be}{\begin{eqnarray}}
\newcommand{\ee}{\end{eqnarray}}

\begin{document}
\title{   Resonating valence bond wavefunctions and classical  interacting dimer models}
\author{Kedar Damle}
\affiliation{\small{Tata Institute of Fundamental Research, 1 Homi Bhabha Road, Mumbai 400005, India}}
\author{Deepak Dhar}
\affiliation{\small{Tata Institute of Fundamental Research, 1 Homi Bhabha Road, Mumbai 400005, India}}
\author{Kabir Ramola}
\affiliation{\small{Tata Institute of Fundamental Research, 1 Homi Bhabha Road, Mumbai 400005, India}}
\begin{abstract} 
We relate properties of nearest-neighbour resonating valence bond (nnRVB) wavefunctions for $SU(g)$ spin
systems on two dimensional bipartite lattices to those of fully-packed classical dimer models with potential energy $V$ on the same lattice.
We define a cluster expansion of $V$ in terms of
$n$-body potentials $V_n$, which are recursively determined from the nnRVB wavefunction on {\em finite subgraphs} of the original lattice. The magnitude of the $n$-body interaction $V_n$ ($n>1$) is of order ${\mathcal O}(g^{-(n-1)})$ for
small $g^{-1}$, while $V_1$ reduces to a constant due to the fully-packed
nature of the model.
At leading non-trivial order on the square lattice, the interacting
dimer model only has two-body interactions $V_2(g)$ that favour two parallel dimers on elementary plaquettes. Setting $g=2$ and using the results of earlier work on this interacting dimer model, we  find that the
long-distance behaviour of the bond-energy correlation function is
dominated by an oscillatory term that decays as $ 1/|\vec{r}|^{\alpha}$ with $\alpha \approx 1.22$ for $SU(2)$ spins. This result is in
remarkable quantitative agreement with earlier direct numerical
studies of the corresponding wavefunction, which give $\alpha \approx 1.20$.

\end{abstract}

\pacs{75.10.Jm}
\vskip2pc

\maketitle
 Spin liquid states of low dimensional insulating magnets, in which
the constituent spins fail to develop any kind of long range order down
to $T=0$ in spite of strong magnetic exchange interactions, are an interesting
consequence of competition between the different magnetic interactions
in the system.  Explicit variational
wavefunctions that encode such behaviours have played
a very important role in our understanding of such states of matter~\cite{Wen_book}. Perhaps the best known examples of such wavefunctions are
the resonating valence bond (RVB) wavefunctions introduced by Anderson and
collaborators~\cite{Fazekas_Anderson,Liang_Doucot_Anderson}.

The simplest of these is the nearest
neighbour RVB (nnRVB) wavefunction $|{\mathbf \Psi}(g)\rangle$ for $SU(g)$
spins on a two dimensional bipartite lattice. It is written as a {\em uniform
amplitude superposition} of {\em all possible} product states
in which each $A$-sublattice spin forms a $SU(g)$ singlet state with one
of its $B$-sublattice neighbours~\cite{footer1}. Although the
$g=2$ wavefunction has been studied on the square lattice for over twenty years now~\cite{Liang_Doucot_Anderson},
we owe a detailed understanding of its properties to much
more recent work~\cite{Albuquerque_Alet,Tang_Henley_Sandvik,Cano_Fendley}:
In Ref~\cite{Albuquerque_Alet,Tang_Henley_Sandvik}, spin and bond-energy correlations were measured in the square lattice case using Monte-Carlo methods to establish that this state
has an exponentially decaying spin correlation function,
$|C_S(\vec{r})| \sim \exp(-|\vec{r}|/\xi)$,
but {\em power-law} bond-energy correlation functions at large $|\vec{r}|$: $|C_{E}(\vec{r})| \sim 1/|\vec{r}|^{\alpha} $, 
with $\alpha \approx 1.20$.  While such a short-ranged spin correlation function is a characteristic property of spin-liquids, the power law form of the bond-energy correlation functions strongly suggests
that the nnRVB wavefunction for $SU(2)$ spins on the square lattice actually describes a critical state on the verge of a transition to an ordered phase in which the bond-energies order.

Here, we develop a precise {\em non-perturbative} mapping that connects
properties of $|{\mathbf \Psi}(g)\rangle$ on a two-dimensional bipartite lattice
to those of a {\em 
classical fully-packed} dimer model on the same lattice, which has
a non-trivial interaction potential $V$ for the dimers in addition to the usual non-overlapping constraint. We define a cluster expansion of $V$ in terms of
$n$-body potentials $V_n$, which are recursively determined from the nnRVB wavefunction on {\em finite subgraphs} of the original lattice. $V_1$ reduces to a constant due to the fully-packed
nature of the dimer model, while the $n$-body interaction $V_n$ ($n>1$) is of order ${\mathcal O}(g^{-(n-1)})$ for
small $g^{-1}$, and thus decreases with $n$. The rate of decrease is controlled by the smallness of $g^{-1}$, which also
controls, via this mapping, the exponential decay of spin correlation functions in $|{\mathbf \Psi}(g)\rangle$. To leading non-trivial order on the square lattice, the interacting
dimer model only has two-body interactions $V_2(g)$ that favour two parallel dimers on elementary plaquettes. Setting $g=2$ and using the results of earlier work~\cite{Alet_PRL,Alet_etal,Papa_Luijten_Fradkin} on this interacting dimer model, we  find that the
long-distance behaviour of the bond-energy correlation function in
$|{\mathbf \Psi}(g=2)\rangle$ is
dominated by an oscillatory term that decays as $ 1/|\vec{r}|^{\alpha}$ with 
\begin{equation}
\alpha \approx 1.22 \; .
\end{equation}

This result is in remarkable agreement with recent studies
of the $SU(2)$ nnRVB wavefunction~\cite{Albuquerque_Alet,Tang_Henley_Sandvik}, and provides a quantitative resolution 
of the surprising coexistence of short-ranged spin
correlations and power-law bond-energy correlations in $|{\mathbf \Psi}(g=2)\rangle$. Our non-perturbative mapping to an interacting fully-packed classical dimer model explains the success of the phenomenological approach of Tang~{\em et. al.}~\cite{Tang_Henley_Sandvik}, who were motivated by
the connection between fully-packed dimer configurations
and {\em ground states} of certain~\cite{Read_Sachdev} large-$N$ $SU(N)$ square lattice antiferromagnets to develop a {\em phenomenological height model} description of the critical aspects of $|{\mathbf \Psi}(g=2)\rangle$,
similar to that used for the long-wavelength properties of
classical fully-packed dimer models~\cite{Fradkin_etal,Henley}.

\begin{figure}
{\includegraphics[width=\hsize]{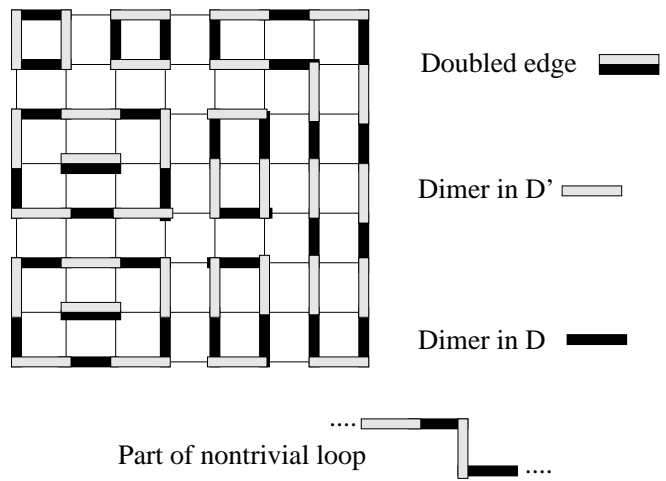}}
\caption{$\langle {\mathcal D}^{'}|{\mathcal D}\rangle_g$ can be represented
in terms of the loop configuration ${\mathcal L}$ obtained from the overlap
loops in the overlap 
diagram generated from dimers in ${\mathcal D}$ and ${\mathcal D}^{'}$. Each overlap loop contributes a factor of $g$ to $\langle {\mathcal D}^{'}|{\mathcal D}\rangle_g$
due to the overlap of the $SU(g)$ singlet states that make up $|{\mathcal D}^{'}\rangle_g$ and $|{\mathcal D}\rangle_g$. By convention,
{\em all} overlap diagrams related by
{\em independent} interchange of black and shaded dimers in {\em each}
overlap loop are identified with the
{\em same loop configuration} ${\mathcal L}$, and therefore $\sum_{{\mathcal D}_1,{\mathcal D}_2}\langle {\mathcal D}_2|{\mathcal D}_1\rangle_g = \sum_{{\mathcal L}}(g)^{n_d({\mathcal L})}(2g)^{n_l({\mathcal L})}$}
\label{basicfigure}
\end{figure}

We begin our analysis by writing
\begin{eqnarray} 
|{\mathbf \Psi}(g)\rangle &= & \sum_{{\mathcal D}}|{\mathcal D}\rangle_g \; \mathrm{, where} \nonumber \\
|{\mathcal D}\rangle_g &=& \prod_{e \in {\mathcal D}} |\Phi_0(g)\rangle_{e} \; .
\end{eqnarray}
Here, the summation is over all fully-packed dimer
configurations ${\mathcal D}$, the product is over all edges $e$ covered
by a dimer in ${\mathcal D}$, and $|\Phi_0(g)\rangle_e$ is the $SU(g)$
singlet state of the two spins connected by edge $e$.
The norm
\begin{equation}
\langle {\mathbf \Psi}(g) | {\mathbf \Psi}(g)\rangle = \sum_{{\mathcal D},{\mathcal D}^{'}} \langle {\mathcal D}^{'} | {\mathcal D}\rangle_g 
\label{ddprimesuperpose}
\end{equation}
can be readily expressed as the
partition function ${\mathcal Z}(g)$ of a classical {\em fully-packed}
 loop model in which each site belongs to exactly one non-intersecting
loop, which can either be a {\em doubled edge} (consisting of a single
edge traversed in both directions) or a {\em non-trivial loop} (consisting
of four or more distinct edges which are each traversed once).
For the case of $SU(g)$ spins, one obtains~\cite{Alet_SUN,Lou_Sandvik_Kawashima}
\begin{eqnarray}
&&{\mathcal Z}_{loop}(g) = \sum_{\mathcal L} w_{loop}(g,{\mathcal L}) \; ,
\label{looppartition}
\end{eqnarray}
with $w_{loop}(g,{\mathcal L}) = (g)^{n_d({\mathcal L})} (2g)^{n_l({\mathcal L})}$, where $n_d({\mathcal L})$ is the number of doubled edges
 and $n_l({\mathcal L})$ the number of non-trivial
loops in the loop configuration
${\mathcal L}$ (Fig~\ref{basicfigure}).
Expectation values and correlators of various physical quantities
can also be represented in terms of estimators that
measure probabilities for various loop-gas configurations. In
the $SU(2)$ case,
the spin correlation function $C_S(\vec{r}) = \langle S(0) \cdot S(\vec{r})\rangle$ is proportional
to the probability that points $0$ and $\vec{r}$ both lie on
the same loop. On the other hand, the connected bond-energy correlation function
$C_{E\mu}(\vec{r}) = \langle B_{\mu}(0) B_{\mu}(\vec{r})\rangle
-\langle B_{\mu}(0)\rangle \langle B_{\mu}(\vec{r})\rangle$
can be related to the probabilities for the
four points $0$, $\hat{\mu}$, $\vec{r}$, and $\vec{r} + \hat{\mu}$
to lie on the same loop or at most on two different loops~\cite{Beach_Sandvik}.
Here, $B_{\mu}(\vec{r}) =(\vec{S}(\vec{r})\cdot \vec{S}(\vec{r}+\hat{\mu}))$
with $\hat{\mu} = \hat{x}, \hat{y}$ representing an elementary lattice translation in the $x$ or $y$ direction.
Since $C_S(\vec{r})$ is proportional to the probability of points $0$ and $\vec{r}$ being on the same loop, the short range nature of $C_S$ implies
that the $g=2$ loop model is in a ``gapped'' phase with predominantly short loops. Indeed, since ${\mathcal Z}_{loop}$ defines a conformally
invariant loop model with a power law distribution of loop sizes for $g=1$~\cite{Kenyon}, we expect that this distribution
becomes exponential for $g>1$. The most natural scenario
is then that the entire $g>1$ short-loop phase
of ${\mathcal Z}_{loop}$ is controlled
(from a renormalization group
standpoint) by the $g= \infty$ fixed point.

Our approach to the existence of power law bond-energy correlation functions in this short loop phase starts with the elementary observation that a loop
gas with extremely short loops can still have long-ranged correlations in the position and orientation of loops.
More precisely, at $g=\infty$, all loops are trivial
in that they have length $2$, and correspond to doubled-edges. The weight of any such loop configuration made up entirely of
doubled edges is always
$g^{N_s/2}$ (where $N_s$ is the number of lattice sites). We now represent
all these doubled-edges by dimers to map such a loop configuration
to a {\em fully-packed dimer
configuration} ${\mathcal D}$ on the same lattice. Thus, loop
configurations that survive the $g \rightarrow \infty$
limit all have equal weight
and define the partition function of a fully-packed dimer model
in which each dimer contributes a factor of $g$ to the weight of
${\mathcal D}$: $w_{dimer}(g, {\mathcal D}) = g^{N_s/2}$ independent of ${\mathcal D}$.

For $g < \infty$, ${\mathcal Z}_{loop}$ also gets
contributions from more general configurations ${\mathcal L}$ consisting
of both doubled-edges and non-trivial loops (with $4$ or
more edges).  Each non-trivial loop in such a finite-$g$ configuration
${\mathcal L}$ can be replaced in exactly two ways by a sequence of doubled-edges on alternating edges of this non-trivial loop.
Thus, a general loop configuration ${\mathcal L}$ with $n_l$ non-trivial loops
 and $n_d$ doubled edges maps to $2^{n_l({\mathcal L})}$ different loop configurations
made up purely of doubled edges, which we represent by dimers. 
Each finite-$g$ configuration ${\mathcal L}$ of the original loop model
thus maps to $2^{n_l({\mathcal L})}$ dimer configurations ${\mathcal D}_{\alpha}$
($\alpha = 1,2 \dots 2^{n_l({\mathcal L})}$).
Next, we distribute $w_{loop}(g,{\mathcal L})$,
the original weight  of ${\mathcal L}$,
 {\em equally} among
these ${\mathcal D}_{\alpha}$.
As a result, each of these $2^{n_l({\mathcal L})}$ different configurations
${\mathcal D}_{\alpha}$ acquire an {\em additional weight} $w(g,{\mathcal L})/2^{n_l({\mathcal L})}$. 

This maps the original loop model to a dimer model
with weights
\begin{eqnarray}
w_{dimer}(g, {\mathcal D}) &=& \sum_{{\mathcal L}|{\mathcal  D}}  \frac{w_{loop}(g,{\mathcal L})}{2^{n_l({\mathcal L})}} 
= \langle {\mathbf \Psi}(g)|{\mathcal D}\rangle_g \; ,
\label{loopweightstodimerweights}
\end{eqnarray}
where ${\mathcal L} | {\mathcal D}$ denotes all loop configurations ${\mathcal L}$ {\em compatible} with the fully-packed dimer cover ${\mathcal D}$, {\em i.e.} obtainable by superposing some fully-packed dimer cover ${\mathcal D}^{'}$
on the given ${\mathcal D}$, and we have used Eqn~(~\ref{ddprimesuperpose}) and (\ref{looppartition}) to obtain the second equality.
The original loop partition function ${\mathcal Z}_{loop}$ is
then equal to the partition sum over all fully-packed
dimer configurations ${\mathcal D}$ with weights $w_{dimer}(g,{\mathcal D})$:
\begin{equation}
{\mathcal Z}_{loop} = {\mathcal Z}_{dimer} = \sum_{{\mathcal D}} w_{dimer}(g,{\mathcal D}) \; .
\end{equation}

As $w_{dimer}(g,{\mathcal D})$ depends on the structure of ${\mathcal D}$, we have provided a precise non-perturbative mapping of the original
loop model to an {\em interacting, classical, fully-packed} dimer model with
potential energy $V(g,{\mathcal D})$ given as
\begin{equation}
V(g,{\mathcal D})= -\log\left(w_{dimer}(g,{\mathcal D})\right) \; .
\end{equation}
We now define a decomposition of the potential energy $V(g,{\mathcal D})$ of a fully packed dimer configuration ${\mathcal D}$ into a sum of $n$-body potential energies $V_n({\mathcal D}_n)$ of subconfigurations ${\mathcal D}_n$
consisting of $n$ distinct dimers from ${\mathcal D}$:
\begin{equation}
V(g,{\mathcal D}) = \sum_n \sum_{{\mathcal D}_n \in {\mathcal D}}V_n({\mathcal D}_n)
\end{equation}

The $V_n$ are determined recursively
from computation of
the weight $w_{dimer}^{{\mathcal G_n}}(g,{\mathcal D}_n)$ of ${\mathcal D}_n$
{\em in the  interacting dimer model on the finite subgraph ${\mathcal G}_n({\mathcal D}_n)$ of the square lattice}. Here, this weight is calculated via Eqn~(\ref{loopweightstodimerweights}) from
the loop model defined on ${\mathcal G}_n({\mathcal D}_n)$,
and the subgraph ${\mathcal G}_n({\mathcal D}_n)$ consists
of the $2n$ vertices covered by dimers of the subconfiguration ${\mathcal D}_n$,
along with all allowed edges between these vertices.

In the first step of this recursive construction, we consider
any particular ${\mathcal D}_1$ and determine the weight
$w_{dimer}^{{\mathcal G}_1}(g,{\mathcal D}_1)$.
The original loop model on ${\mathcal G}_1({\mathcal D}_1)$ has only one valid configuration, which is a doubled edge on the only edge of ${\mathcal G}_1({\mathcal D}_1)$. Using
Eqn~(\ref{loopweightstodimerweights}), we therefore have
$w_{dimer}^{{\mathcal G}_1}(g,{\mathcal D}_1) = g$.
The logarithm of this weight defines the one-body potential
\begin{equation}
V_1\left({\includegraphics[width=0.3cm]{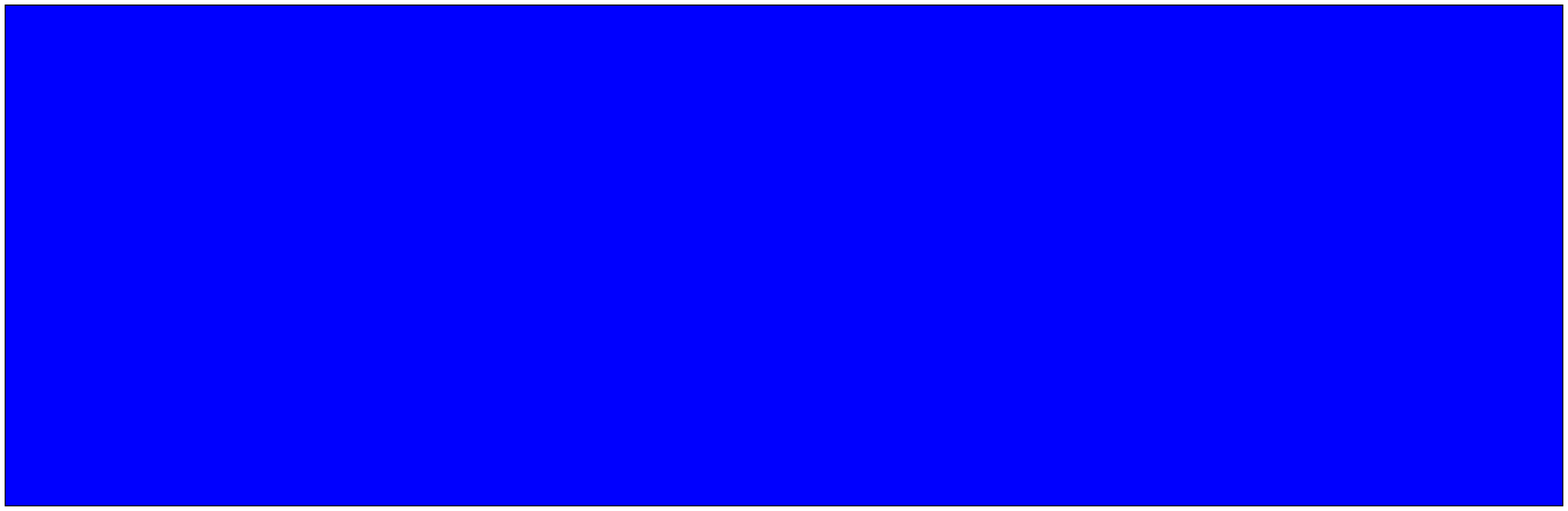}}\right) = V_1\left({\includegraphics[height=0.3cm]{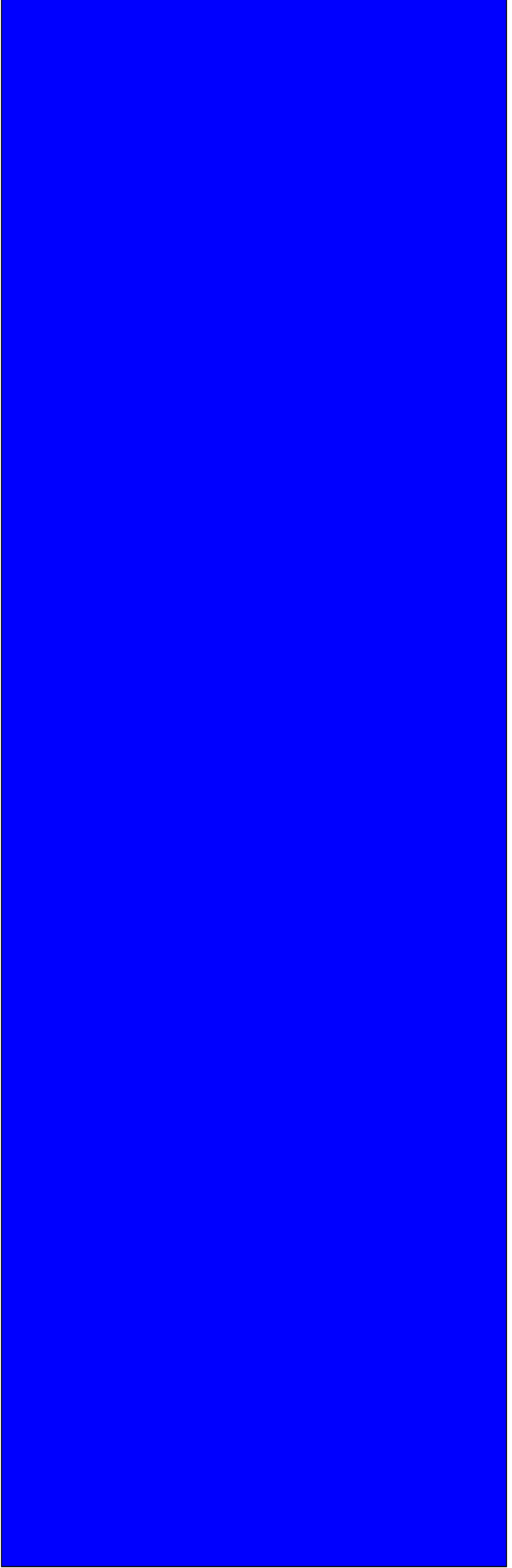}}\right) = -\log(g) \; .
\end{equation}

With this in hand, $V_n$ for arbitrary $n$ can be obtained recursively
from the equation
\begin{equation}
-\log \left[w_{dimer}^{{\mathcal G}_n}(g,{\mathcal D}_n)\right] = V_n({\mathcal D}_n)+\sum_{m=1}^{n-1}
\sum_{{\mathcal D}_m \in {\mathcal D}_n} V_m({\mathcal D}_m) \; ,
\label{basicrecursion}
\end{equation}
where ${\mathcal D}_m \in {\mathcal D}_n$  
denotes all $m$-dimer subconfigurations ${\mathcal D}_m$ of ${\mathcal D}_n$,
and ${\mathcal G}_m({\mathcal D}_m)$ the corresponding subgraphs of
${\mathcal G}_n({\mathcal D_n})$.

For instance, to obtain $V_2$, we consider any particular ${\mathcal D}_2$ 
and determine $w_{dimer}^{{\mathcal G}_2}({\mathcal D}_2)$
as follows: If ${\mathcal G}_2({\mathcal D}_2)$
does {\em not} form a plaquette of the square lattice,
there is only one valid configuration
of the loop model on ${\mathcal G}_2({\mathcal D}_2)$
(consisting of doubled edges that take the place of the dimers in ${\mathcal D}_2$). From
Eqn~(\ref{loopweightstodimerweights}), this gives
$w_{dimer}^{{\mathcal G}_2}({\mathcal D}_2) = g^2$.
On the other hand, when ${\mathcal G}_2({\mathcal D}_2)$ does form
a plaquette of the square lattice, $w_{dimer}^{{\mathcal G}_2}({\mathcal D}_2)$ gets contributions from
two of the three valid configurations of the loop
model on ${\mathcal G}_2({\mathcal D}_2)$. One of these
consists of two doubled edges that take the place of the dimers in ${\mathcal D}_2$, while
the other is a non-trivial loop on the boundary of the plaquette.
Eqn~(\ref{loopweightstodimerweights}) then gives $w_{dimer}^{{\mathcal G}_2}({\mathcal D}_2) = g^2+g$. Knowing $w_{dimer}^{{\mathcal G}_2}({\mathcal D}_2)$,
$V_2$ can be obtained from the recursion relation Eqn~(\ref{basicrecursion}) with $n=2$. Clearly, $V_2$ is non-zero only if the two dimers live on the same plaquette of the square lattice, and in this non-trivial case we obtain
\begin{equation}
V_2\left({\includegraphics[width=0.3cm]{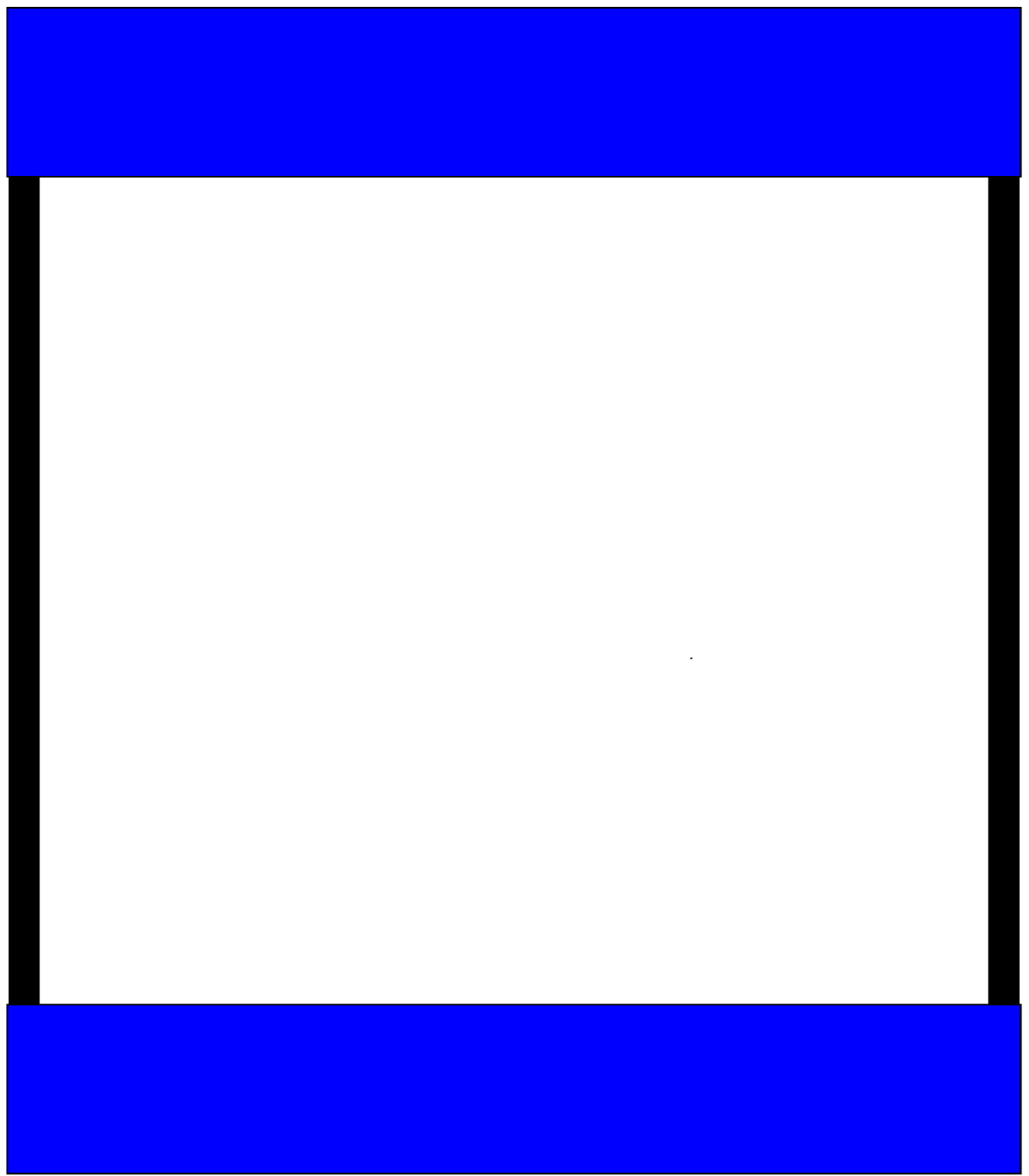}}\right)= V_2\left({\includegraphics[height=0.3cm]{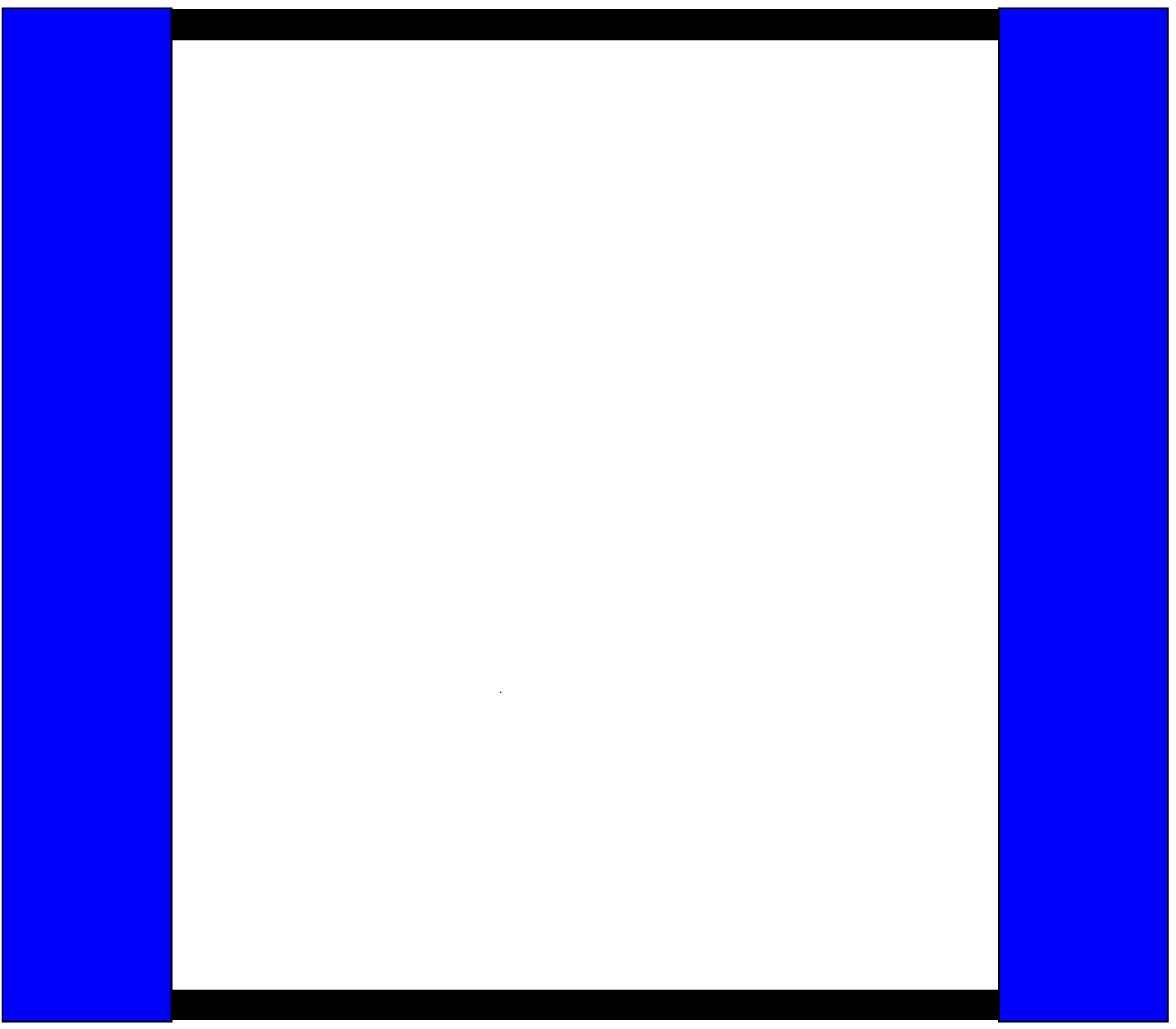}}\right) = -\log(1+g^{-1}) \; .
\end{equation}

Similarly, $V_3$ is seen to be zero unless the three dimers live on a pair of adjacent plaquettes, and in this non-trivial case we obtain
\begin{eqnarray}
&&V_3\left({\includegraphics[width=0.6cm]{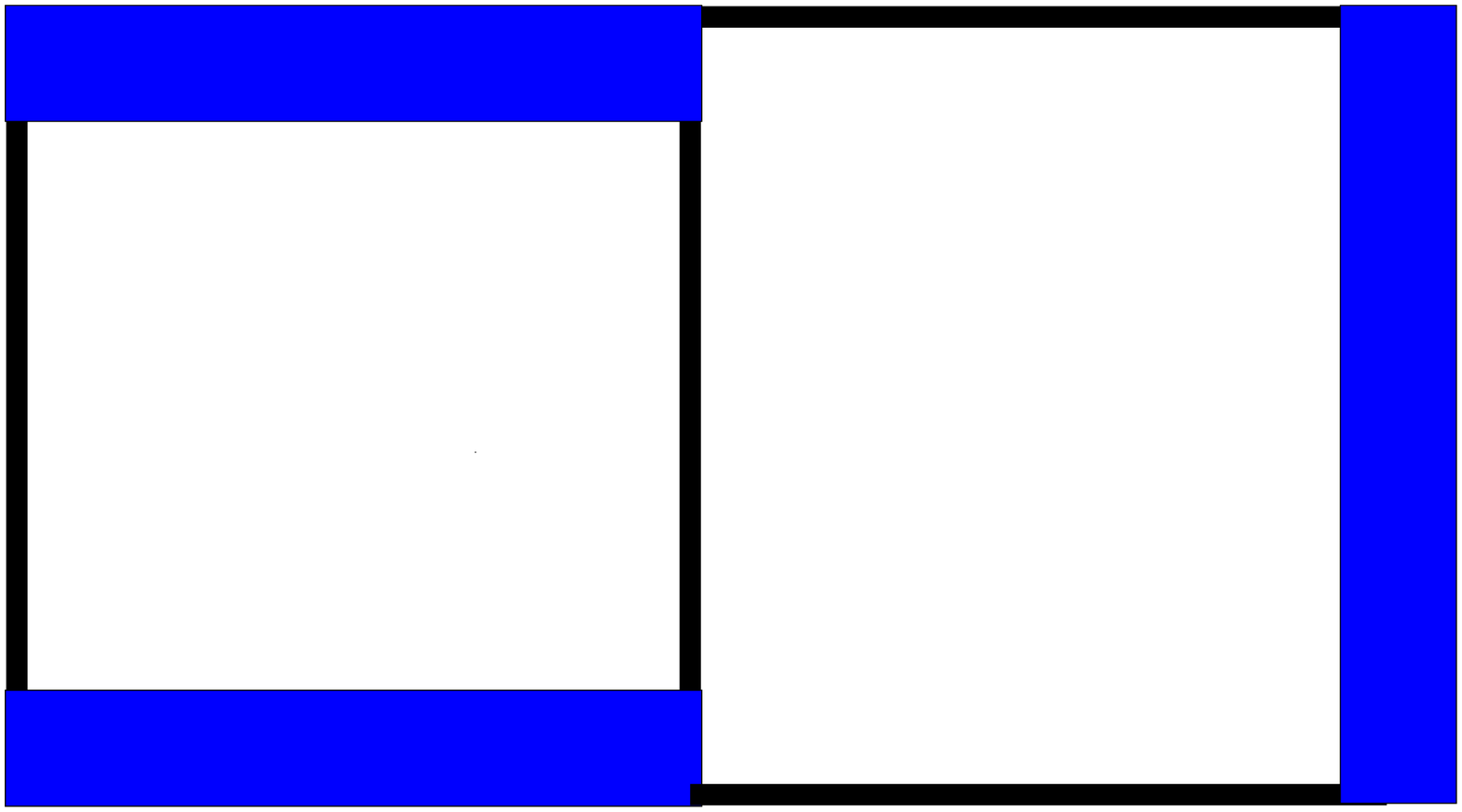}}\right) = -\log\left[(1+g^{-1}+g^{-2})(1+g^{-1})^{-1}\right] \; , \nonumber \\ 
&& V_3\left({\includegraphics[width=0.6cm]{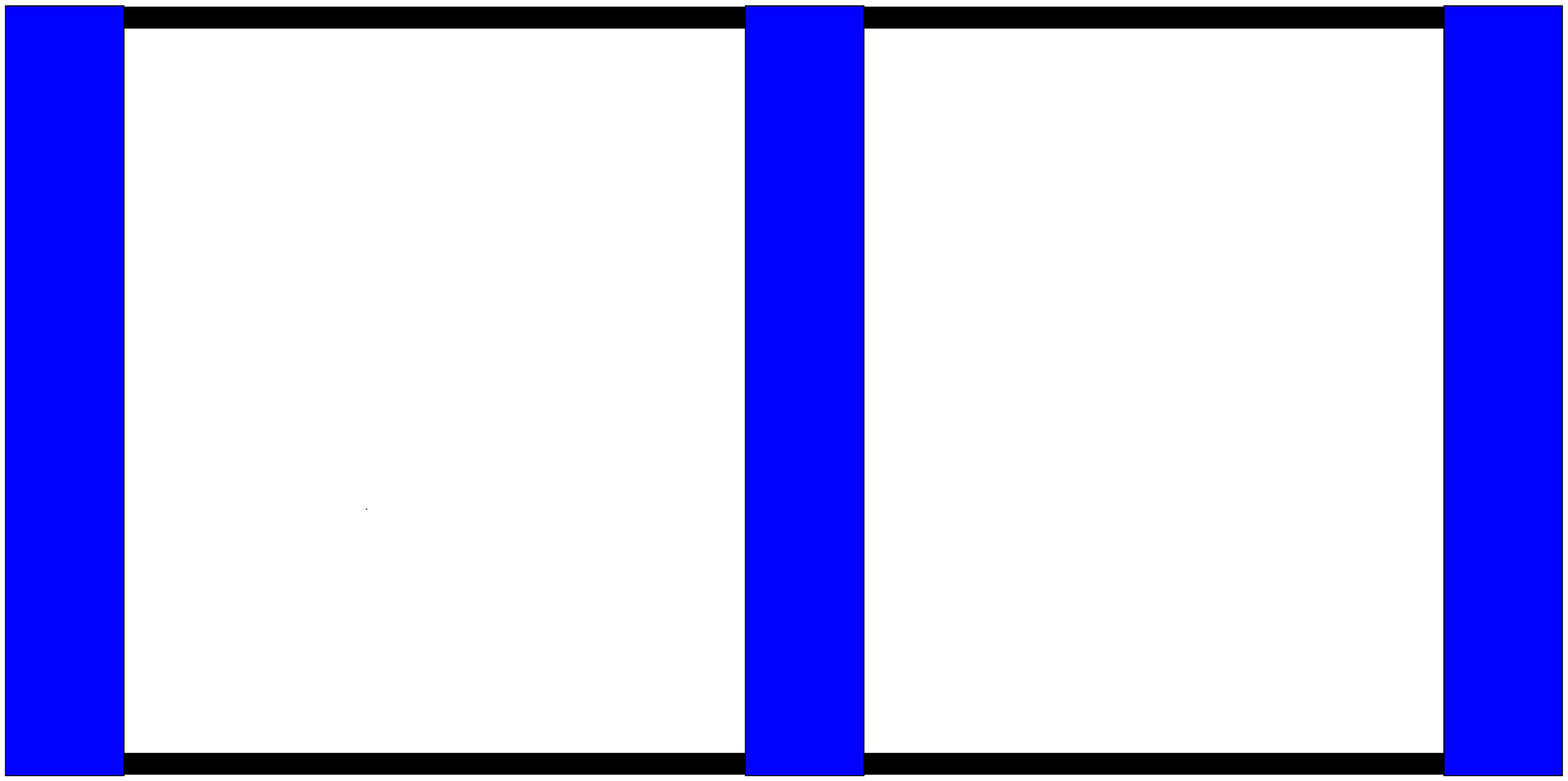}}\right) = -\log\left[(1+2g^{-1})(1+g^{-1})^{-2}\right]  \; ,
\label{v3}
\end{eqnarray}
and the symmetry-related counterparts of Eqn~(\ref{v3}) obtained by using lattice reflection and rotation symmetries.

It is clear that 
each $V_n$ for $n>1$ is of order ${\mathcal O}(g^{-(n-1)})$ when
 the $n$ dimers live on a contiguous set of plaquettes
of the square lattice (sharing edges with each other), and zero otherwise. Our procedure thus expands $V$ in the size of clusters, and is controlled by the smallness of $g^{-1}$. It may be viewed as a classical version of the ``Schrieffer-Wolff''canonical transformation
approach familiar in the context of strongly interacting electronic systems;
in the usual~\cite{Schrieffer_Wolff} Schrieffer-Wolff transformation, the effects of higher energy states in a larger Hilbert space are encoded in
modifications to the effective Hamiltonian that acts in the subspace
of low-energy states, while in our classical
version of this approach, the effects of lower weight configurations
of the full loop model are encoded
in the {\em effective weights} of a restricted subclass
of high-weight {\em dimer model configurations} ${\mathcal D}$.

Expectation values of any operator $\hat{O}$ can be
calculated using the {\em modified estimator} $\bar{{\mathcal P}}_{\hat{O}}({\mathcal D})$ constructed from the original loop gas estimator~\cite{Beach_Sandvik,Alet_SUN} ${\mathcal P}_{\hat{O}}({\mathcal L})$  to correctly encode the effects of the low-weight configurations that are not contained in the dimer model:
\begin{equation}
\bar{{\mathcal P}}_{\hat{O}}({\mathcal D}) = \frac{1}{w_{dimer}(g,{\mathcal D})}\sum_{{\mathcal L}|{\mathcal  D}}  \frac{w_{loop}(g,{\mathcal L}){\mathcal P}_{\hat{O}}({\mathcal L})}{2^{n_l({\mathcal L})}} \; . 
\end{equation}

Consider for instance the spin correlation function. To
zeroth order in $g^{-1}$, the corresponding modified estimator is non-zero only if $0$ and $\vec{r}$ are
connected by a dimer in ${\mathcal D}$, while the ${\mathcal O}(g^{-1})$
correction gives a non-zero result if $0$ and $\vec{r}$
belong to the same {\em flippable plaquette} of ${\mathcal D}$
without being connected by a dimer, and similarly for higher order corrections.
Thus the spin correlation function is expected to decay exponentially
as a function of spatial separation. 
In the case of the bond-energy correlation function $C_{E\mu}(\vec{r})$ defined
earlier, it is readily seen that the leading large $|\vec{r}|$
behaviour is dominated by the first ${\mathcal O}(g^{0})$ term
in the modified estimator, which is obtained by replacing each $B_{\mu}(\vec{r})$ by the dimer occupation number $n_{\mu}(\vec{r})$ of the corresponding
edge in the dimer model:
\begin{equation}
C_{E_{\mu}}(\vec{r}) \sim \langle n_{\mu}(0) n_{\mu}(\vec{r}) \rangle_{V} - \langle n_{\mu}(0) \rangle_{V} \langle n_{\mu}(\vec{r}) \rangle_{V} \; ,
\end{equation}
where $\langle\dots \rangle_{V}$ denotes averages computed in the dimer model with energy $V$.

As is well-known, such fully-packed interacting dimer models admit a microscopic height representation~\cite{Alet_etal,Fradkin_etal,Papa_Luijten_Fradkin,Henley}, which
upon coarse-graining leads to a coarse-grained height action that takes
the form
\begin{eqnarray}
S&=& \pi \rho \int d^2r (\nabla h)^2 + \sum_{p=4,8,12 \dots} y_p\int d^2r \cos(2\pi ph) +\dots \nonumber \\
&&
\end{eqnarray}
where the ellipses denote higher gradient terms and higher powers of
gradients consistent with symmetries~\cite{Fradkin_etal}, and the bare coefficients $\rho$
and $y_p$ are functions of $g$.
The renormalization group theory for this height action is standard~\cite{Jose_etal,Fradkin_etal}: In the present variables, it tells us that there is a line
of fixed points $\rho^{*} = \kappa, y_p^{*}=0$ with $0< \kappa \leq 4$. As long
as the bare values of $\rho$, $y_p$ and the coefficients of
the omitted higher derivative and nonlinear terms are not too large,
the system flows to an attractive fixed point $\kappa(g)$ on
this fixed line. 

As mentioned earlier, the $g \rightarrow \infty$ limit
maps to a non-interacting dimer model since all interactions ($V_n$ with $n>1$) vanish and $V_1$ simply represents the fugacity $g$ of each dimer. Therefore
expect $\kappa(g \rightarrow \infty) = 1/2$ since this is the known value
of the stiffness for a non-interacting dimer model on the square lattice~\cite{Fradkin_etal}. As $g$ is reduced
from $g = \infty$, the leading effect is an interaction $V_2$ that favours flippable
plaquettes. The interacting dimer model with only $V_2$ present
has been studied in detail in Ref~\cite{Alet_PRL,Alet_etal,Papa_Luijten_Fradkin},
which established that the renormalized stiffness $\kappa$
increases monotonically with the magnitude of $V_2$ until
it reaches $\kappa=4$, at which point the system undergoes a Kosterlitz-Thouless like transition to a columnar ordered state. 

To access the physics of the nnRVB wavefunction of $SU(2)$ spins,
we therefore identify the magnitude of $V_2$ with the inverse temperature
$\beta$ of Ref~\cite{Alet_etal}, and set $g=2$ to obtain $\beta = \log(1+g^{-1}) \approx 0.405$~\cite{footer2}.
This places us deep in the high-temperature phase well above the transition to columnar order, and
from Fig. 31 of Ref~\cite{Alet_etal}, we obtain the estimate
$\kappa(g=2) \approx 0.82$.
Calculating the required dimer correlation function from the {\em fixed point} height
action with this value of $\kappa$ using the standard correspondence
between dimer occupation numbers and the height-field~\cite{Fradkin_etal,Henley}, we obtain the leading, large $|\vec{r}|$ form of the bond-energy
correlation functions:
\begin{equation}
C_{E_x}(\vec{r}) \sim \frac{(-1)^x}{|\vec{r}|^{1/\kappa(g)}} \; ; \; C_{E_y}(\vec{r}) \sim \frac{(-1)^y}{|\vec{r}|^{1/\kappa(g)}}
\end{equation}
which gives, upon setting $g=2$, the result advertised earlier. In addition,
both these correlation functions have a subdominant piece which goes as
$(-1)^{x+y}/|\vec{r}|^{2}$ independent of $\kappa(g)$. 

We conclude by noting that our approach generalizes in an obvious
way to include singlets between $A$ and $B$-sublattice sites that
are further apart  on the square lattice, as well as to the $SU(g)$ nnRVB wavefunction on the honeycomb lattice. Again,
we expect the corresponding loop models to be in a short-loop
phase for all $g>1$, and this allows us to access the physics of these
wavefunctions for $g \geq 2$ by mapping to an interacting dimer model;
on the honeycomb lattice, the leading interaction terms will favour {\em flippable hexagons}, while the presence of longer range singlets will introduce
additional interactions on the square lattice. Possible generalizations to various three-dimensional bipartite
lattices are more intriguing, since it is not obvious
that the loop model will be in a short-loop phase for all $g>1$ on any
of these lattices.

\textit{Acknowledgements}
We acknowledge computational resources of TIFR and research
support from the Indo-French Centre for the Promotion of Advanced Research (IFCPAR/CEFIPRA) under Project 4504-1 (KD), and from the Indian DST
via grants DST-SR/S2/RJN-25/2006 (KD) and DST-SR/S2/JCB-24/2005 (DD).
We thank F.~Alet for a critical reading of an earlier draft of this manuscript. 
One of us (KD) would also like to thank  the organizers
and all participants of the Toulouse Worskhop on Quantum Magnetism, particularly F.~Alet, R. Moessner, and A. Sandvik,
for interesting and insightful discussions, and thank ICTS-TIFR
and IISc Bangalore for hospitality during completion of this work.


\begin{thebibliography}{999}

\bibitem{Wen_book} {\em Quantum Field Theory of Many-Body Systems}, X.~G.~Wen,
Oxford University Press (2004).

\bibitem{Fazekas_Anderson} P.~Fazekas and P.~W.~Anderson, Philosophical Magazine {\bf 30}, 423-440 (1974).

\bibitem{Liang_Doucot_Anderson} S.~Liang, B.~Doucot, and P.~W.~Anderson,
Phys. Rev. Lett. {\bf 61}, 365 (1988).

\bibitem{footer1} Spins on the $A$ sublattice carry the fundamental
representation of $SU(g)$ while those on the $B$ sublattice carry
the complex conjugate of this.



\bibitem{Albuquerque_Alet} A.~F.~Albuquerque and F.~Alet, Phys. Rev. B
{\bf 82}, 180408(R) (2010).

\bibitem{Tang_Henley_Sandvik} Y.~Tang, A.~W.~Sandvik, and C.~L.~Henley, 
Phys. Rev. B {\bf 84}, 174427 (2011).

\bibitem{Cano_Fendley} J.~Cano and P.~Fendley, Phys. Rev. Lett. {\bf 105},
067205 (2010).

\bibitem{Alet_PRL} F.~Alet {\it et. al.}, Phys. Rev. Lett. {\bf 94}, 235702 (2005).

\bibitem{Alet_etal} F.~Alet {\it et. al.}, Phys. Rev. E {\bf 74}, 041124 (2006).

\bibitem{Papa_Luijten_Fradkin} S.~Papanikolaou, E.~Luijten, and E.~Fradkin,
Phys. Rev. B {\bf 76}, 134514 (2007).


\bibitem{Read_Sachdev} N.~Read and S.~Sachdev, Phys. Rev. B {\bf 42},
4568 (1990).

\bibitem{Fradkin_etal} E.~Fradkin {\em et. al.}, Phys. Rev B {\bf 69},
224415 (2004).

\bibitem{Henley} C.~L.~Henley, arXiv:cond-mat/0311345v1 (2003).


\bibitem{Alet_SUN}  K.S.D. Beach, F. Alet, M. Mambrini and S. Capponi, Phys. Rev. B {\bf 80}, 184401 (2009).

\bibitem{Lou_Sandvik_Kawashima} J.~Lou, A.W.~Sandvik, and N.~Kawashima, Phys. Rev. B {\bf 80}, 180414 (2009). 

\bibitem{Beach_Sandvik} K. S. D. Beach and A. W. Sandvik, Nuclear Physics B 750, 142 (2006).
 



\bibitem{Kenyon}  R.~Kenyon, arXiv:1105.4158 (unpublished).


\bibitem{Schrieffer_Wolff} J.~R.~Schrieffer and P.~A.~Wolff, Phys. Rev. {\bf 149}, 491 (1966).




\bibitem{Jose_etal} J.~V.~Jose, L.~P.~Kadanoff, S.~Kirkpatrick, and
D.~R.~Nelson, Phys. Rev. B {\bf 16}, 1217 (1977).

\bibitem{footer2} It may be possible to improve this estimate
of $\beta$ by partial resummation of the effects of 3-body and higher couplings,
along the lines of the analysis in Ref.~{\protect{\cite{David_Matthieu}}}.

\bibitem{David_Matthieu} D.~Schwandt, M.~Mambrini, and D.~Poilblanc,
Phys. Rev. B {\bf 81}, 214413 (2010).

\end{thebibliography}
\end{document}